\documentclass[onecollarge,natbib]{svjour2}
\bibpunct{[}{]}{;}{n}{}{,} 
\smartqed  
\usepackage{graphicx}
\usepackage{mathptmx}      
\usepackage{latexsym}
\usepackage{amsmath}
\usepackage{amssymb}
\newcommand{\Real}{\mathop{\rm Re}}
\newcommand{\Img}{\mathop{\rm Im}}
\newcommand{\C}{{\if mm {{\rm C}\mkern -15mu{\phantom{\rm t}\vrule}}
\mkern +10mu \else \leavemode \hbox{I}\kern -.17em \hbox{C} \fi}}

\journalname{}
\begin{document}

\title{Ultracold Scattering and Universal Correlations\thanks{
Based on a talk presented at the 22nd
European Conference on Few-Body Problems in Physics (September
9--13, 2013, Cracow, Poland). The paper is to be published in
\textit{Few-Body Systems}. This work was supported in part by the Heisenberg-Landau Program}}

\titlerunning{Ultracold Scattering and Universal Correlations}

\author{E.A. Kolganova}

\authorrunning{E.A. Kolganova}

\institute{E.A. Kolganova \at
              Bogoliubov Laboratory of Theoretical Physics\\
							Joint Institute for Nuclear Research\\
							Joliot-Curie 6,141980 Dubna, Moscow region, Russia\\
               \email{kea@theor.jinr.ru}           
}


\maketitle

\begin{abstract}
Particle-dimer scattering below and above the three-body threshold is studied using Faddeev differential equations.
Correllations between the observables are shown and some analogies between three-nucleon and three-atom systems are dicussed.

\keywords{Three-body systems \and Faddeev equations \and scattering and resonances }
\end{abstract}


\medskip

Ultracold scattering of different atoms and universal 
correlations have recently received considerable attention, in both experiment and theory. 
A remarkable linear correlation between the triton binding energy and the doublet nucleon - dueteron scattering length
was first pointed out by Phillips \cite{Phillips}, and is known as  Phillips line. A similar approximate linear correlation  between the binding energy of the helium trimer and the $^4$He-dimer scattering length can be observed in the three-atomic system~\cite{Roudnev}.
The Phillips line has not been explained for many years and only two decades later Efimov and Tkachenko
 \cite{EfimovT} showed that the approximate linearity of the Phillips line is due to the fact that 
 the interval covered by the calculations with various NN-potentials is not wide enough.
Roudnev and Cavagnero in \cite{Roudnev} suggested a modified Phillips line with a dimensionless variables
\begin{equation}
\label{mPh}
\alpha=\ell_{\rm sc} \sqrt{-2\mu \epsilon_d}/\hbar \, , \quad \omega = 1/\sqrt{\frac{E_3}{\epsilon_d} - 1},
\end{equation}
where $\alpha$ is a dimensionless scattering length and $\omega$ is a dimensionless parameter characterizing the three-body binding energy; $\epsilon_d$ and  $E_3$  define the energies of the two- and three-body bound states, respectively; $\mu$ is the reduced mass of the particle-dimer system, and $\ell_{\rm sc}$ is the particle-dimer scattering length.
Their idea to use such parametrization was based on the paper by Efimov and Tkachenko \cite{EfimovT} who noted that for a
weakly bound three-body state the approximate correlation $\epsilon_d - E_3 \approx \hbar^2/(2\mu \ell_{\rm sc})$
should hold.  
The universal properties of systems with a large scattering length can well be described by an effective field theory. 
As shown in \cite{Braaten}, a zero-range model formulated in field theoretical terms is able to simulate the scattering situation.
Aplications of the effective field theory to systems we deal with below could be found in recent review articles \cite{Platter,Hammer}.

Here we study this problem using the Faddeev formalism~\cite{FM} in the hard core model~\cite{JPhysB}.
 In our approach,  in order to change the value of the scattering length, we 
modify the interatomic potential by
multiplying the original interatomic HFD-B potential from~\cite{Aziz} 
by the strength factor $\lambda$: $V (x) = \lambda V_{HFD-B} (x)$.
This strategy is extensively used in literature~\cite{Esry, YaF99,Kievsky13}  
because  it allows one to deal with a wide range of scattering lengths.

Following Refs.\cite{YaF99,review}, we use the Faddeev formalism~\cite{FM}.  
After the partial-wave and angular analysis the Faddeev  differential equations read
\begin{equation}
\label{FadPart}
   \left[-\displaystyle\frac{\partial^2}{\partial x^2}
            -\displaystyle\frac{\partial^2}{\partial y^2}
            +l(l+1)\left(\displaystyle\frac{1}{x^2}
            +\displaystyle\frac{1}{y^2}\right)
    -E\right]\Phi_l(x,y)=\left\{
            \begin{array}{cl} -V(x)\Psi_l(x,y), & x>c \\
                    0,                  & x<c\,.
\end{array}\right.
\end{equation}
Here $x$ and $y$ stand for the standard Jacobi variables, $c$ for the core range, and $l$ for partial angular momentum.
$V(x)$ is the He-He central potential acting outside the core domain. The partial
wave function $\Psi_l(x,y)$ is related to the Faddeev components $\Phi_l(x,y)$ (see details in \cite{JPhysB}).

The functions $\Phi_{l}(x,y)$ satisfy the boundary conditions
\begin{equation}
\label{BCStandard}
      \Phi_{l}(x,y)\left.\right|_{x=0}
			=\Phi_{l}(x,y)\left.\right|_{y=0}=0\, , \quad  \Psi_{l}(c,y) =0\,.
\end{equation}
In addition, for the bound state problem, the partial wave Faddeev
components $\Phi_{l}(x,y)$ should satisfy the asymptotic boundary condition given by
\begin{equation}
\label{AsBCPart}
      \Phi_l(x,y)  = 
      \delta_{l0}\psi_d(x)\exp({\rm i}\sqrt{E_3-\epsilon_d} y)
      \left[{\rm a}_0+o\left(y^{-1/2}\right)\right] 
  + \displaystyle\frac{\exp({\rm i}\sqrt{E_3}\rho)}{\sqrt{\rho}}
                \left[A_l(\theta)+o\left(\rho^{-1/2}\right)\right],
\end{equation}
while for the problem of two-fragment scattering
states,  the asymptotic boundary condition for the function $\Phi_{l}(x,y)$ reads
\begin{equation}
\label{AsBCPartS}
    \begin{array}{rcl}
      \Phi_l(x,y;p) & = &
      \delta_{l0}\psi_d(x)\left\{\sin(py) + \exp({\rm i}py)
      \left[{\rm a}_0(p)+o\left(y^{-1/2}\right)\right]\right\} \\
      && +
  \displaystyle\frac{\exp({\rm i}\sqrt{E}\rho)}{\sqrt{\rho}}
                \left[A_l(\theta,p)+o\left(\rho^{-1/2}\right)\right].
    \end{array}
\end{equation}
Here $\psi_d(x)$ denotes  the dimer wave function, $E$
stands  for the scattering energy given by
$E=\epsilon_d+p^2$ with $\epsilon_d$ being the dimer energy,
and $p$ corresponds to the relative momentum conjugate to the
variable $y$. The variables $\rho=\sqrt{x^2+y^2}$ and
$\theta=\arctan\displaystyle\frac{y}{x}$ are the
hyperradius and hyperangle, respectively. 
One of the important parameters, which determines the behavior of asymptotic boundary conditions  (5), is
the coefficient ${\rm a}_0(p)$ that is the elastic scattering amplitude. The functions $A_l(p,\theta)$
provide us, at $E>0$, with the corresponding partial-wave Faddeev breakup amplitudes. The scattering
length $\ell_{\rm sc}$ is given by
\begin{equation}
\label{sclen}
\ell_{\rm sc}=-\displaystyle\frac{\sqrt{3}}{2}\,
\begin{array}{c}\phantom{a}\\
{\rm lim}\,\\
\mbox{\scriptsize$p\rightarrow0$}
\end{array}\,
\frac{{\rm a}_0(p)}{p}.
\end{equation}
\begin{figure}
\centering
  \includegraphics[width=0.46\textwidth]{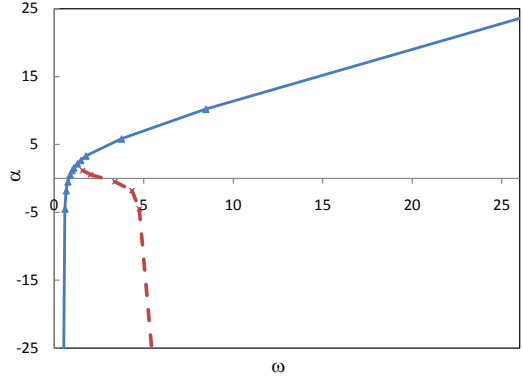}
\caption{The modified Phillips line plotted with the rescaled dimensionless parameters (\ref{mPh}) for the HFD-B potential~\cite{Aziz}.
Solid line and dashed line correspond  to the binding energy of the helium trimer and the virtual state energy, respectively.}
\label{fig:1}       
\end{figure}

Now let us discuss the results obtained by the solution of the Faddeev integro-differential equations (\ref{FadPart})-(\ref{AsBCPartS}).
In Fig. 1 we plot the dependence of  the dimensionless correlation $\alpha$ on $\omega$ computed 
for the HFD-B potential where the scattering length varies from -7000 \AA \ to 8500 \AA.
Almost linear relation between the rescaled parameters $\alpha$ and $\omega$ is held 
for large values of $\omega$,
for which a weakly bound three-body state exists (solid curve in Fig.1). 
The nonlinear regime is observed when the scaled scattering length $\alpha$ is small.
This  nonlinearity originates from the presence of the near-threshold virtual states shown by the dashed curve line in Fig.1. This virtual state violates the linear dependence $\alpha$($\omega$)  at the small scattering length. When the scattering length  is large and negative, the virtual state turns into a bound state. We studied this mechanism in detail in~\cite{YaF99}.  
In this limit of the large but negative scattering length the linear dependence is again restored. We note that similar results for the weakly bound three-body state were obtained in Ref.\cite{Roudnev} where  
in order to demonstrate the universality of the linear dependence $\alpha( \omega)$ the curve was plotted for different two-body potentials. 


Recently, Kievsky  and Gattobigio \cite{Kievsky13}  investigated the universal behavior  in the elastic scattering below the dimer breakup threshold calculating the $^4$He atom-dimer effective-range function. They showed, that the same parametrization of the effective-range function also described the nucleon-deutron scattering below the deutron breakup threshold. 

The similarity of properties of the atomic helium trimer and nuclear $nnp$ systems can also be seen in the S-matrix formalism.
Let us compare the results obtained  by the S-matrix formalism for atomic helium trimer and nuclear $nnp$ system. 
To compute the s-state  $nnp$ and $^4$He$_3$  resonances, we use the algorithm described in Refs.~\cite{YaF99,YaF97}. We note that the resonances lying on the two-body unphysical sheet are the poles of the S-matrix.
The unphysical sheet is connected with the physical one by crossing the spectral interval $(\epsilon_d,\, 0)$ between the two-body (deuteron or dimer) energy $z\!  =\! \epsilon_d$ and breakup threshold $z\! =\! 0$. Therefore, as it was analytically proved in Ref.~\cite{Motovilov}, these resonances are the roots of the S-matrix on the physical sheet. 
Using this fact, we solve  two--dimensional Faddeev integro--differential equations (\ref{FadPart}) with the 
boundary conditions (\ref{BCStandard}) and (\ref{AsBCPartS}) at complex energies $z$ 
and extract the truncated s-state scattering matrix 
\begin{equation}
\label{S0}
{\rm S}_0(z)=1+2{\rm i}{\rm a}_0(z)\,.
\end{equation}

We start our analysis by considering the $^4$He$_3$ system in the only state with the
angular momentum $l=0$. As the interatomic He-He-interaction we employ 
the semiempirical potential HFD-B~\cite{Aziz}. We assumed $\hbar^2/{\rm m} = 12.12$\,K\AA$^2$
where ${\rm m}$ stands for the mass of the $^4{\rm He}$ atom. Notice that for the
HFD-B potential the $^4$He-dimer energy is -1.685\,mK \cite{review}.

In this case, we study graph surfaces of the real and imaginary
parts of the scattering matrix ${\rm S}_0(z)$ in the domain of
its holomorphy (curve 1 on Fig.2, see for details~\cite{YaF99}).
The root lines of the functions $\Real{\rm S}_0(z)$ and $\Img{\rm S}_0(z)$ 
presented in Fig.~2a in the dimensionless case were obtained for
 the grid parameters \mbox{$N_\theta=N_\rho=600$} and
\mbox{$\rho_{\rm max}=600$}\,{\AA}. The resonances are associated with the intersection 
points of the curves  $\Real{\rm S}_0(z) =0$ and $\Img{\rm S}_0(z)=0$ .

%
\begin{figure*}
\centering
\vspace*{15px}
\includegraphics[width=0.49\textwidth]{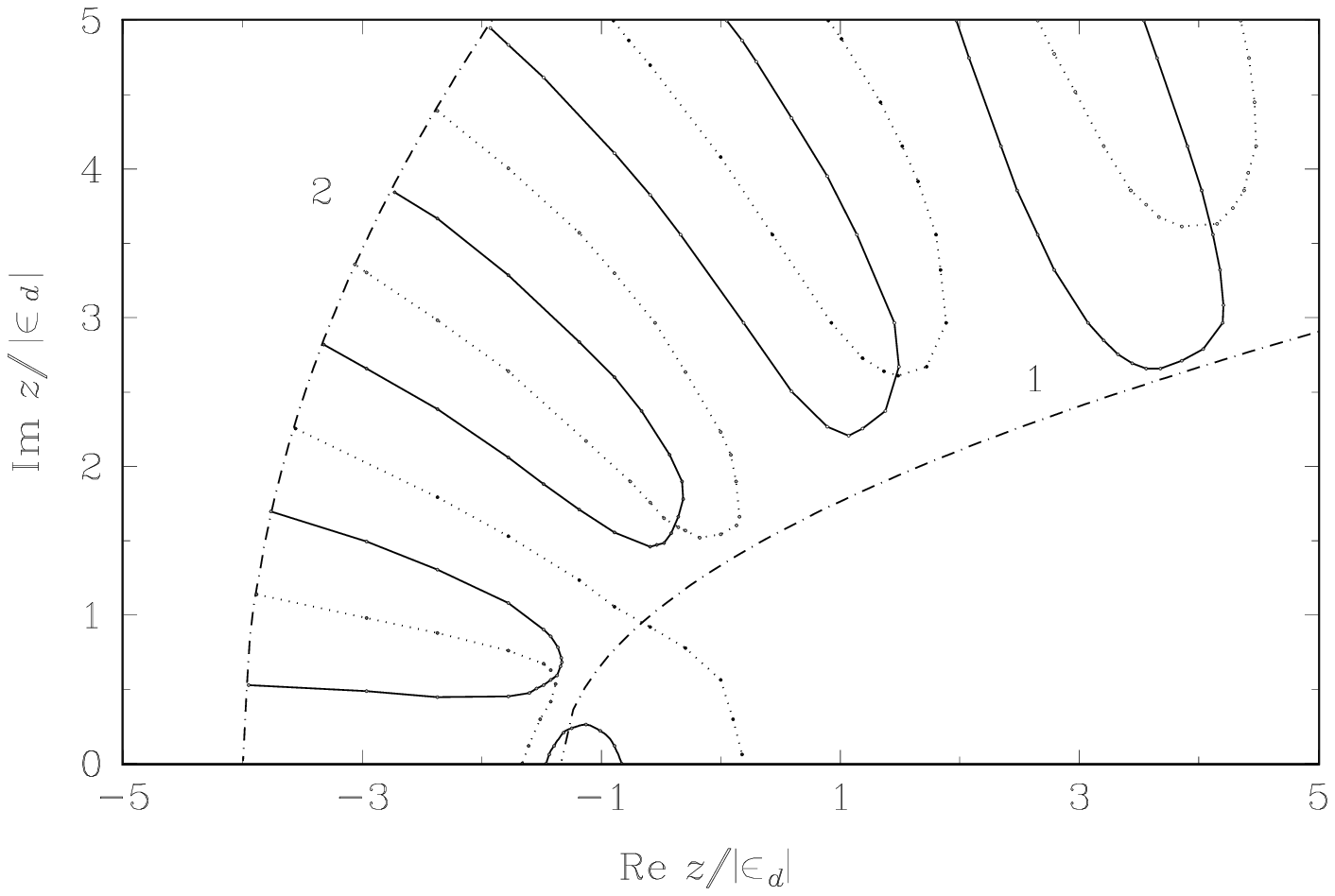}
\includegraphics[width=0.49\textwidth]{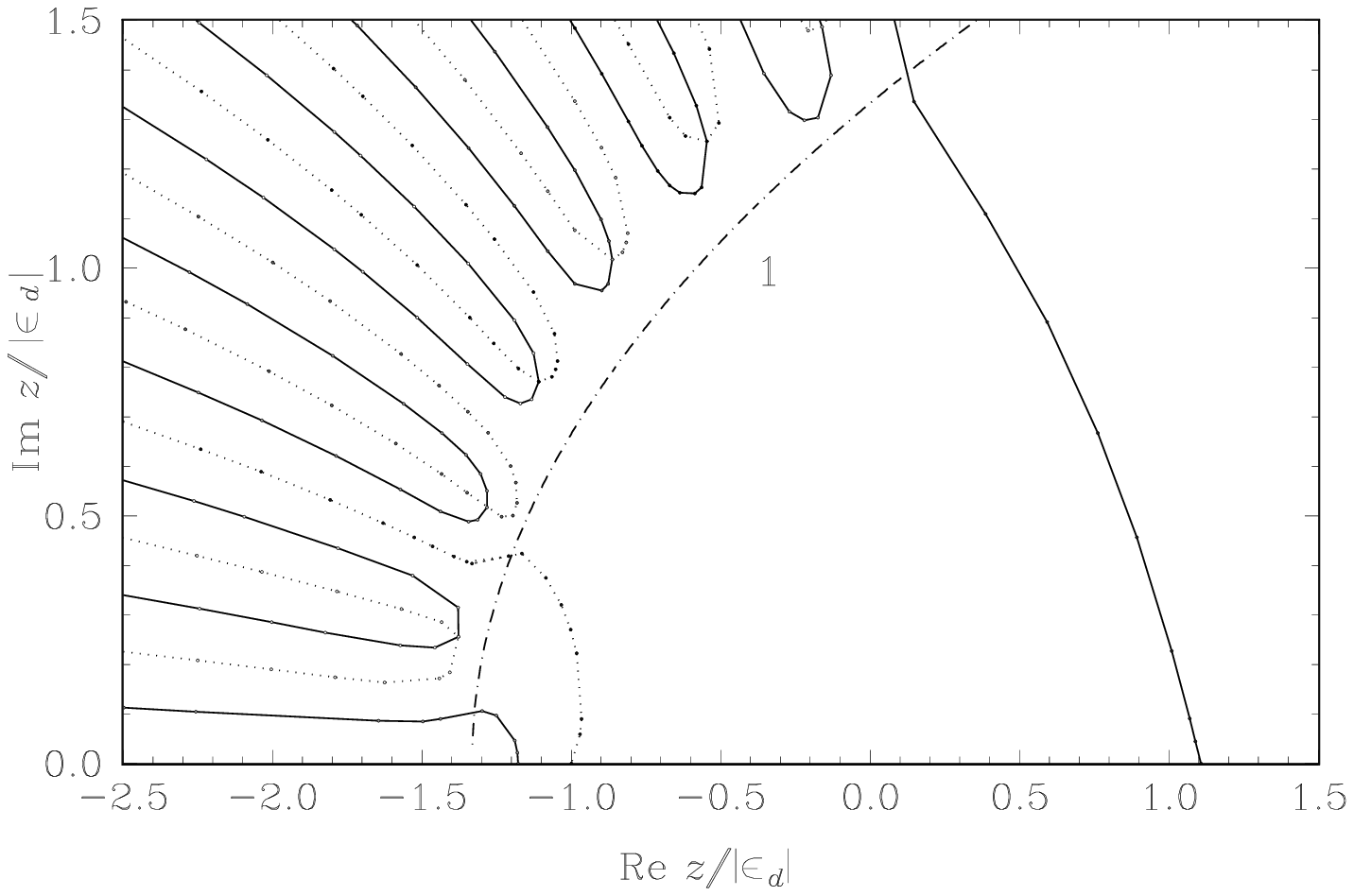}
\vspace*{-130px}
\hspace{1cm}a)\hspace{8cm} b)
\caption{Root locus curves of the real and imaginary parts of
the scattering matrix ${\rm S}_0(z)$ in the case of helium trimer (a) and
$nnp$ system (b). The solid lines correspond
to $\Real {\rm S}_0(z)=0$ while the tiny dashed lines to $\Img{\rm S}_0(z)=0$. 
Curves 1 and 2 denote the boundaries of the domains where the scattering matrix ${\rm S}_0(z)$ and
Faddeev components  $\Phi_l(x,y;z)$ can be analytically continued in $z$, respectively.}
\label{fig:2}       
\end{figure*}

Next we study the $nnp$ system using the Malfliet--Tjon potential MT~I--III~\cite{MT} as NN-interaction.
Contrary to the previous case, this system is a fermionic one, nevertheless, it is described by the set of equations 
similar to  (\ref{FadPart}) - (\ref{AsBCPartS}). The only difference is that for the doublet $nd$ scattering the function 
$\Phi_l(x,y)$ is the two-component vector. Therefore, we have to solve the system of two coupled integro-differential equations. 
The details of calculation and the procedure how to compute the $nnp$ system can be found in Refs. \cite{FM,YaF97}.

In this case,  we have succeeded to find one root $z_{\rm res}\! =\!
E_{\rm T}^v$ of the function $S_{0}(z)$ corresponding to the known
virtual state of the $nnp$ system at the total spin S=1/2.  For the grid described by
\mbox{$N_\theta=N_\rho=1400$}, {$\rho_{max}=100 $}~fm  we
have found $z_{\rm res}\! =\! -2.699$~MeV. 
We see that it is situated $0.475$~MeV to the left from the $nd$ threshold $\epsilon_d\! =\! -2.224$~MeV (in the
MT~I--III model \cite{MT}).  Our result agrees well with the shift $\epsilon_d\! -\! z_{\rm res} =0.48$~MeV
found from experimental data on $nd$ scattering ~\cite{TUNL}. 
Moreover, this value is in good agreement with the previous theoretical results known from the literature~\cite{OrlovT,OrlovN,NP2001}:
the value $0.48$ MeV was obtained in ~\cite{OrlovT} as a pole of the T-matrix with a separable version of the MT potential;
in~\cite{OrlovN} the authors studied an effective-range function for the doublet $nd$ scattering while in~\cite{NP2001} 
the complex scaling method was used.

Concerning the structure of the S-matrix in the $nnp$ system, we also studied the root locus
lines of the real and imaginary parts of the scattering matrix ${\rm S}_0(z)$ obtained for the
grid parameters \mbox{$N_\theta=N_\rho=240$} and
\mbox{$\rho_{\rm max}=100$}\, fm (see Fig.2b). 
Qualitatively, the structure of the S-matrix for two very different systems presented in Fig.2 a) and b) is very similar. 
This finding strongly supports the results  by Kievsky and Gattobigio \cite{Kievsky13}, and  also connects  the universal 
behavior of atomic systems with the large two-body scattering length to the known physics of the nuclear systems 
with the near-threshold virtual state.

We have investigated theoretically the particle-dimer scattering below  and above the three-body threshold. We considered two systems 
-- atomic helium trimer and nuclear $nnp$ systems -- and showed that despite the different nature of these systems they show 
universal  correlation. This universality is captured by using Faddeev differential equations and studying the S-matrix structure.

\begin{acknowledgements}
The author is grateful  to A.K.Motovilov and W.Sandhas for their interest in this work and constant support.
\end{acknowledgements}



\end{document}